\documentclass[graybox,envcountchap]{svmult}
\pdfoutput=1
\usepackage{graphicx}
\usepackage[sectionbib]{chapterbib}

\usepackage{multicol}        
\usepackage[bottom]{footmisc}
\usepackage{enumitem}

\usepackage{newtxtext}       %

\usepackage{url}
\usepackage{pifont}

\usepackage{algorithm}
\usepackage{algpseudocode}
\usepackage{booktabs}
\usepackage{multirow}
\usepackage{wrapfig}
\usepackage{listings}
\usepackage{siunitx}
\usepackage{pgfplotstable}
\usepackage{array}
\usepackage{subcaption}
\usepackage{newfloat}
\usepackage{hyphenat}

\usepackage{pdflscape}
\usepackage{xspace}

\usepackage{diagbox} 

\usepackage[disable]{todonotes}

\newcommand*{\eg}{e.g.,\@\xspace}
\newcommand*{\etal}{\emph{et~al.}\@\xspace}
\newcommand\YAMLcolonstyle{\color{red}\mdseries}
\newcommand\YAMLkeystyle{\color{black}\bfseries}
\newcommand\YAMLvaluestyle{\color{blue}\mdseries}

\makeatletter

\newcommand\language@yaml{yaml}

\expandafter\expandafter\expandafter\lstdefinelanguage
\expandafter{\language@yaml}
{
  keywords={true,false,null,y,n},
  keywordstyle=\color{darkgray}\bfseries,
  basicstyle=\YAMLkeystyle,                                 
  sensitive=false,
  comment=[l]{\#},
  morecomment=[s]{/*}{*/},
  commentstyle=\color{purple}\ttfamily,
  stringstyle=\YAMLvaluestyle\ttfamily,
  moredelim=[l][\color{orange}]{\&},
  moredelim=[l][\color{magenta}]{*},
  moredelim=**[il][\YAMLcolonstyle{:}\YAMLvaluestyle]{:},   
  morestring=[b]',
  morestring=[b]",
  literate =    {---}{{\ProcessThreeDashes}}3
                {>}{{\textcolor{red}\textgreater}}1     
                {|}{{\textcolor{red}\textbar}}1 
                {\ -\ }{{\mdseries\ -\ }}3,
}

\lst@AddToHook{EveryLine}{\ifx\lst@language\language@yaml\YAMLkeystyle\fi}
\makeatother

\newcommand\ProcessThreeDashes{\llap{\color{cyan}\mdseries-{-}-}}

\usepackage{wasysym}

\DeclareFloatingEnvironment[
    fileext=los,
    listname={List of Listings},
    name=Listing,
    placement=!h,
    within=none,
]{listing}

\lstdefinestyle{docker}{
  language=bash,
  morekeywords={RUN,FROM,MAINTAINER},
  showstringspaces=false,
  frame=none,
}
\lstdefinelanguage{ansible}{
  morekeywords={name,vars,hosts,tasks,roles,role},
  keywordstyle=\bfseries,
  morecomment=[l][\textit]\#,
  morecomment=[s][\bfseries]{\{\{}{\}\}},
}
\lstdefinestyle{ansible}{
  language=ansible,
  basicstyle=\scriptsize\ttfamily,
}

\def\postbreak{%
  \raisebox{0ex}[0ex][0ex]{\ensuremath{\hookrightarrow\space}}}

\lstdefinestyle{searchstringstyle}{
	basicstyle=\ttfamily\footnotesize,
	breakatwhitespace=false,         
	breaklines=true,                 
	captionpos=t,                    
	keepspaces=true,                 
	numbers=none,                    
	numbersep=5pt,                  
	showspaces=false,                
	showstringspaces=false,
	showtabs=false,                  
	tabsize=2,
	frame=single
}

\definecolor{mauve}{rgb}{0.58,0,0.82}
\definecolor{dkgreen}{rgb}{0,0.6,0}
\definecolor{gray}{rgb}{0.5,0.5,0.5}

\definecolor{pblue}{rgb}{0.13,0.13,1}
\definecolor{pgreen}{rgb}{0,0.5,0}
\definecolor{pred}{rgb}{0.9,0,0}
\definecolor{pgrey}{rgb}{0.46,0.45,0.48}

\lstdefinelanguage{JavaScript}{
  keywords={break, case, catch, continue, debugger, default, delete, do, else, false, finally, for, function, if, in, instanceof, new, null, return, switch, this, throw, true, try, typeof, var, void, while, with},
  morecomment=[l]{//},
  morecomment=[s]{/*}{*/},
  morestring=[b]',
  morestring=[b]",
  ndkeywords={class, export, boolean, throw, implements, import, this},
  sensitive=true
}

\lstdefinestyle{JavaStyle}{%
  language=Java,
  showspaces=false,
  showtabs=false,
  breaklines=true,
  showstringspaces=false,
  breakatwhitespace=true,
  commentstyle=\color{pgreen},
  keywordstyle=\color{pblue},
  stringstyle=\color{pred},
  columns=flexible,
  basicstyle={\small\ttfamily},
  numbers=left,
  frame=tb,
  xleftmargin=2em,
  framexleftmargin=2em
}

\lstdefinestyle{JavaScriptStyle}{%
  language=JavaScript,
  showspaces=false,
  showtabs=false,
  breaklines=true,
  showstringspaces=false,
  breakatwhitespace=true,
  commentstyle=\color{pgreen},
  keywordstyle=\color{pblue},
  stringstyle=\color{pred},
  columns=flexible,
  basicstyle={\small\ttfamily},
  numbers=left,
  frame=tb,
  xleftmargin=2em,
  framexleftmargin=2em
}

\usepackage{tikz} 
\usetikzlibrary{shadows,calc}

\colorlet{innercolor}{black!60}
\colorlet{outercolor}{gray!05}

\pgfdeclarelayer{shadow} 
\pgfsetlayers{shadow,main}

\begin{document}
\frontmatter
\mainmatter


\title{The Software Heritage Open Science Ecosystem}
\author{Roberto Di Cosmo \and Stefano Zacchiroli}
\institute{%
  Roberto Di Cosmo \at Inria and Universit\'e Paris Cit\'e, France, \email{roberto@dicosmo.org}
  \and
  Stefano Zacchiroli \at LTCI, T\'el\'ecom Paris, Institut Polytechnique de Paris, France \email{stefano.zacchiroli@telecom-paris.fr}
}
\maketitle
\label{SWH:ch}

\def\swhMetricsDate{February 2023}

\def\swhAbstractText{

  Software Heritage~\cite{swhcacm2018} is the largest public archive of software
  source code and associated development history, as captured by modern version
  control systems. As of \swhMetricsDate\ it has archived more than 12 billion
  unique source code files and 2 billion commits, coming from more than 180
  million collaborative development projects. In this chapter we describe the
  Software Heritage ecosystem, focusing on research and open science use cases.

  On the one hand Software Heritage supports empirical research on software by
  materialising in a single Merkle direct acyclic graph the development history
  of public code~\cite{swhipres2017}. This giant graph of source code artifacts
  (files, directories, and commits) can be used---and has been used---to study
  repository forks~\cite{pietri:msr:2020}, open source
  contributors~\cite{ieee-sw-gender-swh, msr-2022-foss-licenses,
    icse-seis-2022-gender}, vulnerability propagation, software provenance
  tracking~\cite{swh-provenance-emse}, source code indexing, and more.

  On the other hand Software Heritage ensures availability and guarantees
  integrity of the source code of software artifacts used in any field that
  relies on software to conduct experiments, contributing to making research
  reproducible. The source code used in scientific experiments can be
  archived---e.g., via integration with open access
  repositories~\cite{swh-hal-02475835}---referenced using persistent
  identifiers~\cite{swhipres2018} that allow downstream integrity checks, and
  linked to/from other scholarly digital artifacts~\cite{swh-ICMS2020}.

}

\abstract*{\swhAbstractText}
\abstract{\swhAbstractText}

\newpage

\section{The Software Heritage Archive}
\label{SWH:sec:archive}

Software Heritage~\cite{swhipres2017,swhcacm2018} is a non-profit initiative
started by Inria in partnership with UNESCO to build a long-term universal
archive specifically designed for software source code, capable of storing
source code files and directories, together with their full development
histories.

Software Heritage's mission is to collect, preserve, and make easily accessible
the source code of \emph{all publicly available software}, addressing the needs
of a plurality of stakeholders, ranging from cultural heritage to public
administrations and from research to industry.

\begin{figure}[ht]
\centering
\includegraphics[width=\linewidth]{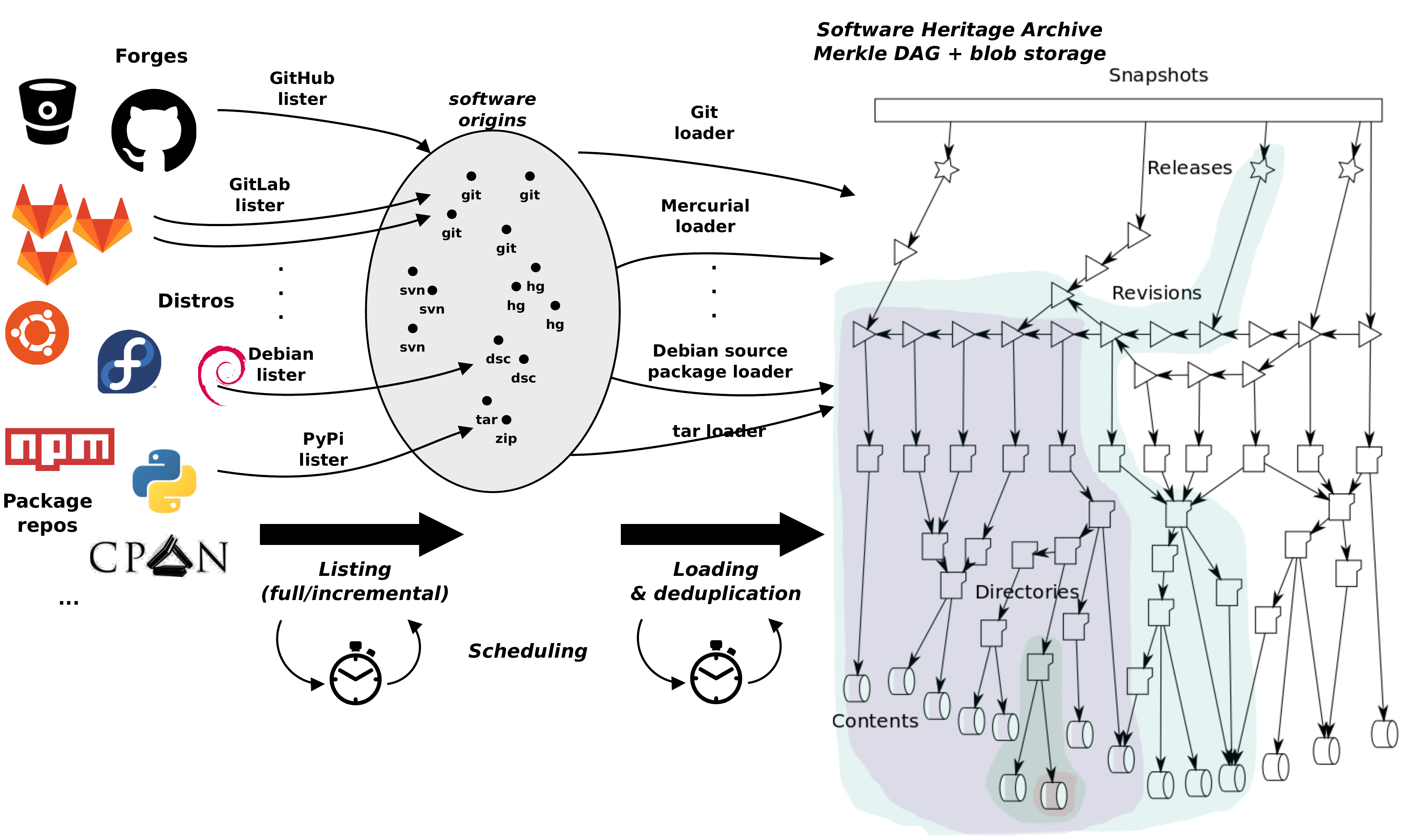}
\caption{\label{swh:fig:swhcrawler}%
  Software Heritage data flow: crawling (on the left) and archival (right)}
\end{figure}

The key principles that underpin this initiative are described in detail in two
articles written for a broader audience in the early
years of the project~\cite{swhipres2017,swhcacm2018}. One of these principles was to avoid
any \emph{a priori} selection of the contents of the archive, to avoid the risk
of missing relevant source code, whose value will only become apparent later on.
Hence one of the strategies enacted for collecting source code to archive
is the large-scale automated crawling of major software development forges and
distributions, as shown in Figure~\ref{swh:fig:swhcrawler}.

As a consequence of this automated harvesting, there is no guarantee that the
content of the archive only contains quality source code, or only code that
builds properly: curation of the contents will need to happen at a later stage,
via human or automated processes that build a view of the archive for specific
needs. It may also happen that the archive ends up containing content that needs
to be removed, and this required the creation of a process to handle take down
requests following current legal regulations.\footnote{See \url{https://www.softwareheritage.org/legal/content-policy/} for details.}

The sustainability plan is based on several pillars. The first one is the
support of Inria, a national research institution that is involved for the long
term. A second one is the fact that Software Heritage provides a common
infrastructure catering to the needs of a variety of stakeholders, ranging from
industry to academia, from cultural heritage to public administrations. As a
consequence, funding comes from a diverse group of sponsors, ranging from IT
companies to public institutions.
Finally, an extra layer of archival security is provided by a network of
independent international mirrors that maintain each a full copy of the archive.\footnote{More details can be found at
  \url{https://www.softwareheritage.org/support/sponsors} and
  \url{https://www.softwareheritage.org/mirrors}.}

We recall here a few key properties that set Software Heritage apart from other
scholarly infrastructures:
\begin{itemize}

\item Software Heritage \emph{proactively} archives \emph{all software}, making it possible to
  store and reference any piece of publicly available software relevant to a
  research result, independently from any specific field of endeavour, and even
  when the author(s) did not take any step to have it
  archived~\cite{swhipres2017, swhcacm2018};

\item Software Heritage stores source code with its development history in a uniform data
  structure, a Merkle Directed Acyclic Graph (DAG)~\cite{Merkle}, that allows to provide uniform,
  \emph{intrinsic} identifiers for tens of billions archived software
  artifacts, independently of the version control system (VCS) or package
  distribution technology used by software developers~\cite{cise-2020-doi}.

\end{itemize}

\smallskip\noindent\textbf{Relevance for software ecosystems.}
Software Heritage relates to software ecosystems, according to the seminal
definition of Messerschmitt \etal~\cite{messerschmitt2003software} in two main ways.
On the one hand, software products are associated to source code artifacts that
are versioned and stored in VCSs.  For Free/Open Source Software (FOSS), and
more generally public code, those artifacts are distributed publicly and can be
mined to pursue various goals.  Software Heritage collects and preserves
observable artifacts that originates from open source ecosystems, enabling
others to access and exploit them in the foreseeable future.

On the other hand, Software Heritage provides the means to foster the sharing
of even more of those artifacts in the specific case of open scientific
practices---what we refer to as the ``open science ecosystem'' in this chapter.
Contrary to software-only ecosystems, the open science ecosystem encompasses a
variety of software and non-software artifacts (e.g., data, publications);
Software Heritage has contributed to this ecosystem the missing piece of
long-term archival and referencing of scientifically-relevant software source
code artifacts.

\subsection{Data Model}
\label{swh:sec:datamodel}

Modern software development produces multiple kinds of source code artifacts
(\eg source code files, directories, commits), which are usually stored and
tracked in version control systems, distributed as packages in various formats,
or otherwise. 

When designing a software source code archive that stores source code with its
version control history coming from a disparate set of platforms, there are
different design options available. One option is to keep a verbatim copy of all
the harvested content, which makes it easy to immediately reuse the package or
version control tool. However, this approach can result in storage explosion: as
a consequence of both social coding practices on collaborative development
platforms and the liberal licensing terms of open source software, those source
code artifacts end up being massively duplicated across code hosting and
distribution platforms.

Choosing a data structure that minimizes duplication is better for long-term
preservation and the ability to identify easily code reuse and duplication.

This is the choice made by Software Heritage. Its data model is a Direct
Acyclic Graph (DAG) that leverages classical ideas from content addressable
storage and Merkle trees~\cite{Merkle}, that we recall briefly here.

\begin{figure}[ht]
  \centering
  \includegraphics[width=0.7\textwidth]{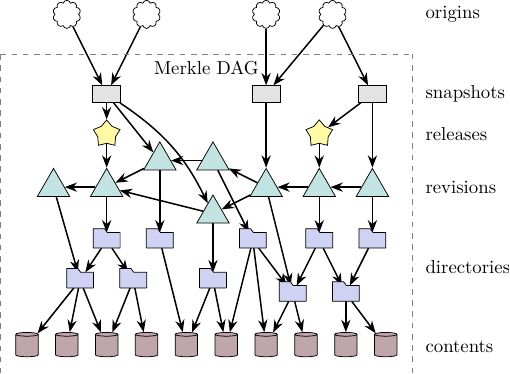}
  \caption{Data model of the Software Heritage archive: a directed acyclic
    graph (DAG) linking together deduplicated software artifacts shared across
    the entire body of (archived) public code}%
  \label{swh:fig:datamodel}
\end{figure}

As shown in Figure~\ref{swh:fig:datamodel}, the Software Heritage DAG is
organized in five logical layers, which we describe below from bottom to top.

\textbf{Contents} (or ``blobs'') form the graph's leaves, and contain the raw
content of source code files, not including their filenames (which are
context-dependent and stored only as part of directory entries).

\textbf{Directories} are associative lists mapping names to directory entries
and associated metadata (e.g., permissions). Each entry can point to content
objects (``file entries''), revisions (``revision entries'', e.g., to represent
git submodules or subversion externals), or other directories (``directory
entries'').

\textbf{Revisions} (or ``commits'') are point-in-time representations of the
entire source tree of a development project. Each revision points to the root
directory of the project source tree, and a list of its parent revisions (if
any).

\textbf{Releases} (or ``tags'') are revisions that have been marked by
developers as noteworthy with a specific, usually mnemonic, name (e.g., a
version number like ``4.2''). Each release points to a revision and might
include additional metadata such as a changelog message, digital signature,
etc.

\textbf{Snapshots} are point-in-time captures of the full state of a project
development repository. While revisions capture the state of a single
development line (or ``branch''), snapshots capture the state of \emph{all}
branches in a repository and allow to reconstruct the full state of a
repository that has been deleted or modified destructively (e.g., rewriting its
history with tools like ``git rebase'').

\textbf{Origins} represent the places where artifacts have been encountered in
the wild (e.g., a public Git repository) and link those places to snapshot
nodes and associated metadata (e.g., the timestamp at which crawling happened),
allowing to start archive traversals pointing into the Merkle DAG.

The Software Heritage archive is hence a giant graph containing nodes
corresponding to all these artifacts and links between them as graph edges.

What makes this DAG capable of deduplicating identical content is the fact that
each node is identified by a cryptographic hash that concisely represent its
contents, and that is used in the SWHID identifier detailed in the next
section. For the blobs that are the leafs of the graph, this identifier is just
a hash of the blob itself, so even if the same file content can be present in
multiple projects, its identifier will be the same, and it will be stored in the
archive only once, like in classical content addressable
storage~\cite{swh-venti-cas}.  For internal nodes, the identifier is computed
from the aggregation of the identifiers of its children, following the
construction originally introduced by Ralph Merkle~\cite{Merkle}: as a
consequence, if a same directory, possibly containing thousands of files, is
duplicated across multiple project, its identifier will stay the same, and it
will be stored only once in the archive. The same goes for revision, releases
and snapshots.

\begin{figure}[ht]
  \centering
  \includegraphics[width=\textwidth]{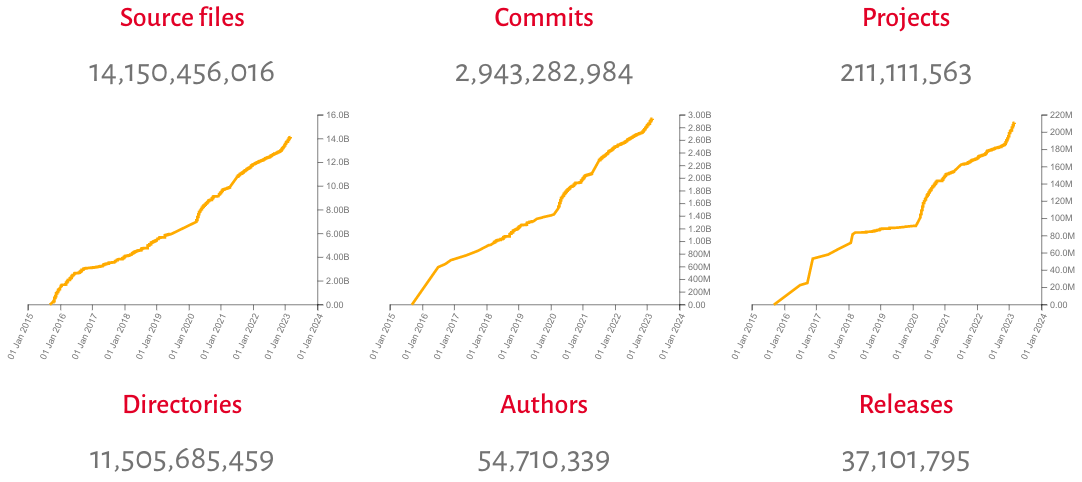}
  \caption{\label{swh:fig:growth} Evolution of the Software Heritage Archive
    over time (\swhMetricsDate)}
\end{figure}

In terms of size the archive grows steadily over time as new source code
artifacts get added to it, as shown in Figure~\ref{swh:fig:growth}. As of
\swhMetricsDate, the Software Heritage archive contained over 14 billions
unique source code files, harvested from more than 210 million software
origins.\footnote{See \url{https://archive.softwareheritage.org} for these and
  other up-to-date statistics.}

\subsection{Software Heritage Persistent Identifiers (SWHIDs)}
\label{swh:sec:swhid}

\begin{figure}[htbp]
\centering
\includegraphics[width=\linewidth]{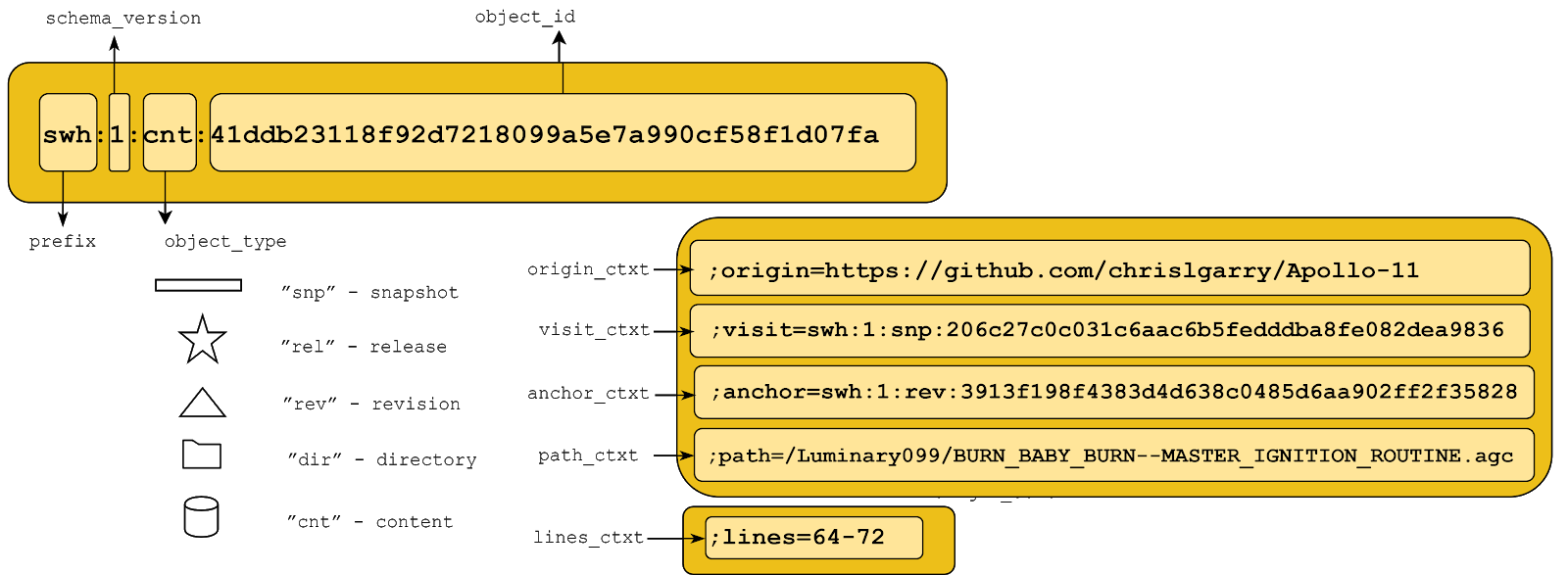}
\caption{\label{swh:fig:swhid}
Schema of the Software Heritage identifiers (SWHID)}
\end{figure}

As part of the archival process, a \emph{Software Heritage Persistent
  Identifier (SWHID)}, is computed for each source code artifact added to the
archive and can be used later to reference, lookup, and retrieve it from the
archive. The general syntax of SWHIDs is shown in Figure~\ref{swh:fig:swhid}.\footnote{See
  \url{https://docs.softwareheritage.org/devel/swh-model/persistent-identifiers.html}
  for the full specification of SWHIDs.}

SWHIDs are URIs~\cite{rfc3986} with a simple syntax. \emph{Core} SWHIDs start
with the \texttt{"swh"} URI scheme; the colon (\verb|:|) is used as separator
between the logical parts of identifiers; the schema version (currently
\verb|1|) is the current version of this identifier schema; then follows the
type of source code artifacts identified; and finally comes a hex-encoded
(using lowercase ASCII characters) cryptographic signature of this object,
computed in a standard way, as detailed in~\cite{swhipres2018, cise-2020-doi}.

Core SWHIDs can then be complemented by \emph{qualifiers} that carry contextual
\emph{extrinsic} information about the referenced source code artifact:
\begin{description}
\item[\itshape origin:] the \emph{software origin} where an object has been found or observed in the wild, as an URI;
\item[\itshape visit:] persistent identifier of a \emph{snapshot} corresponding to a specific \emph{visit} of a repository containing the designated object;
\item[\itshape anchor:] a \emph{designated node} in the Merkle DAG relative to which a \emph{path to the object} is specified;
\item[\itshape path:] the \emph{absolute file path}, from the \emph{root directory} associated to the \emph{anchor node}, to the object;
\item[\itshape lines:] \emph{line number(s)} of interest, usually pointing within a source code file.
\end{description}

The combination of core SWHIDs and qualifiers provides a powerful means of
referring in a research article all source code artefacts of interest.

\bigskip

By keeping all the development history in a single global Merkle DAG.
Software Heritage offers unique opportunities for \emph{massive analysis
  of the software development landscape}. By archiving and referencing all the
publicly available source code, the archive also constitutes the ideal place to
\emph{preserve research software artifacts} and offers powerful mechanisms to
\emph{enhance research articles} with precise references to relevant fragments
of source code, and contributes an essential building block to the software
pillar of Open Science.

\section{Large Open Datasets for Empirical Software Engineering}
\label{SWH:sec:ese}

The availability of large amounts of source code that came with the growing
adoption of open source and collaborative development has attracted the interest
of software engineering researchers since the beginning of the 2000's, and
opened the way to large-scale empirical software engineering studies and a
dedicated conference, Mining Software Repositories.

Several shared concerns emerged over time in this area, and we recall here
some of the ones that are relevant for the discussion in this chapter.

One issue is the significant overhead involved in the systematic extraction of
relevant data from the publicly available repositories and their analysis for
testing research hypotheses. Building a very large scale dataset containing
massive amounts of source code with its version control history is a complex
undertaking and requires significant resources, as shown in seminal work by
Mockus in 2009~\cite{mockus2009amassing}.  The lack of a common infrastructure
spawned a proliferation of ad hoc pipelines for collecting and organising source
code with its version control history, a duplication of efforts that were
subtracted to the time available to perform the intended research and hindered
their reusability. A few initiatives were born with the intention of improving
this unsatisfactory state of affairs: Boa~\cite{swh:dyer2013boa} provides
selected datasets (the largest and most recent one at the time of writing consists
of about 8 million GitHub repositories sampled in October 2019) and a dedicated domain
specific language to perform efficient queries on them, while World of
Code~\cite{mockus2019woc} collects git repositories on a large scale and
mainains dedicated data structures that ease their analysis.

The complexity of addressing the variety of existing code hosting platforms and
version control systems resulted in focusing only on subsets of the most popular
ones, in particular the GitHub forge and the git version control system, which
raises another issue: the risk of introducing bias in the results.
In empirical sciences, \emph{selection bias}~\cite{heckman-1990-selection-bias}
is the bias that originates from performing an experiment on a
non-representative subset of the entire population under study. It is a
methodological issue that can lead to threats to the \emph{external validity}
of experiments, i.e., incorrectly concluding that the obtained results are
valid for the entire population, whereas they might only apply to the selected
subset. In empirical software engineering a common pattern that could result in
selection bias is performing experiments on software artifacts coming from a
relatively small set of development projects. It can be mitigated by ensuring
that the project set is representative of the larger set of projects of
interest, but doing so could be challenging.

Finally, there is the issue of enabling \emph{reproducibility} of large-scale
experiments---i.e., the ability to replicate the findings of a previous
scientific experiment, by the same or a different team of scientists, reusing
varying amounts of the artifacts used in the original
experiment~\cite{ivie-2018-reproducibility-survey}.\footnote{For the sake of
conciseness we do not differentiate here between repeatability, reproducibility,
and replicability; we refer instead the interested reader to the ACM terminology
available at
\url{https://www.acm.org/publications/policies/artifact-review-and-badging-current}.
To varying degrees Software Heritage helps with all of them, specifically when
it comes to mitigating the risk of losing availability to source code
artifacts.}
Large-scale empirical experiments in software engineering might easily require
shipping \emph{hundreds} of GiB up to a few TiB of source code artifacts as
part of replication packages, whereas current scientific platform for data self
archival usually cap at tens of GiB.\footnote{For comparison: the total size of source
code archived at Software Heritage is $\approx$1\,PiB at the time of writing.}

The comprehensiveness of the Software Heritage archive, that makes available the
largest public corpus of source code artifacts in a single logical place, helps
with all these issues:

\begin{itemize}
\item reduces the \emph{opportunity cost} of conducting large-scale
experiments by offering at regular intervals as \emph{open datasets} full dumps of the
archive content
\item contributes to \emph{mitigate} selection bias and the associated external validity threats by providing a corpus that strives to be \emph{comprehensive} for researchers conducting empirical software engineering experiments targeting large project populations.
\item the persistence offered by an independent digital archive, run by a non profit open organisation, eases the
process of ensuring the \emph{reproducibility} of large-scale
experiments, avoiding the need to re-archive the same open source code artifacts in
multiple papers, a wasteful practice that should be avoided if possible.
Using Software Heritage it is enough to thoroughly document in replication packages the SWHIDs
(see Section~\ref{swh:sec:swhid}) of all source code artifacts\footnote{As it will become clear in Section~\ref{swh:sec:swhid}, in most cases it will be sufficient to list the SWHIDs of the releases or repository snapshots.} used in an
empirical experiment to enable other scientists to reproduce the experiments
later on~\cite{swh-ICMS2020}.
\end{itemize}

Table~\ref{swh:esecomp} summarises the above points, comparing with a few other
infrastructures designed specifically for software engineering studies.

\begin{table}[!t]
\centering
\begin{center}
\caption{Comparison of infrastructures for performing empirical software engineering research}
\label{swh:esecomp}
\begin{tabular}{c|cc|c|c}
\diagbox{Criteria}{Infrastructure}
                                 & SWH      & SWH graph    & Boa           & World of Code   \\
                                 &  on S3             & (on premise) &               &                 \\  
\toprule
host organisation                & \multicolumn{2}{c|}{non profit foundation} & research project & research project \\
&&&\\
purpose       & \multicolumn{2}{c|}{archival \& research}     &               &                 \\
&&&\\
scope                            & \multicolumn{2}{c|}{all platforms}      & GitHub, SourceForce & Git hosting \\
&&&\\
dataset                          & \multicolumn{2}{c|}{open}         & closed        & closed          \\
&&&\\
access                           & \multicolumn{2}{c|}{free}         & on demand     & on demand       \\
&&&\\
query language                   & SQL Athena    & graph API    & custom DSL    & custom API      \\
&&&\\
cost                             & 5\$/TB        & ~10K\$ setup & free          & free            \\
&&&\\
dataset update frequency                   & \multicolumn{2}{c|}{6 months} & $\approx$ yearly &  $\approx$ yearly \\
&&&\\
\multirow{2}{*}{reproducibility} & \multicolumn{2}{c|}{named dataset} & named dataset & named dataset \\
                                 & \multicolumn{2}{c|}{SWHID list}   &               &                 \\
\bottomrule
\end{tabular}
\end{center}
\end{table}

\bigskip
In the rest of this section we briefly describe the datasets that Software
Heritage curates and maintains to the benefit of other researchers in the field
of empirical software engineering.

Before detailing the available datasets, we recall that building and maintaining
the Software Heritage infrastructure that is instrumental to build them is a
multimillion dollars undertaking. We are making significant efforts to reduce
the burden on the prospective users, by providing dumps at regular intervals
that help with reproducibility and making them directly available on public
clouds like AWS. Researchers can then either run their queries directly
on the cloud, paying only the compute time, or download them for exploiting
them on their own infrastructure.

To give an idea of the associated costs for researchers, SQL queries on the
graph datasets described in ~\ref{swh:graphdataset} can be performed using
Amazon Athena for approximately 5\$ per Terabyte scanned at the time of writing.
For example, an SQL query to get the 4 topmost commit verb stems from over 2 billion
revisions scans approximately 100 Gigabytes of data, and provides the user with
the answer in less than a minute, for a total cost of approximately 50 cents,
a minimal fraction of the cost one would incur to set up an on premise solution.

When SQL queries are not enough (typically when a graph traversal is needed),
the cost of a cloud solution may quickly become significant, and it may become
more interesting to set up an on premise solution. The full compressed graph
dataset can be exploited using medium range server grade machines that are
accessible for less than 10 thousand dollars.

\subsection{The Software Heritage Datasets}
The entire content of the Software Heritage archive is publicly available to
researchers interested in conducting empirical experiments on it. At the
simplest level, the content of the archive can be browsed interactively using
the Web user interface at \url{https://archive.softwareheritage.org/} and
accessed programmatically using the Web API documented at
\url{https://archive.softwareheritage.org/api/}. These access paths, however,
are not really suitable for large-scale experiments due to protocol overheads
and rate limitations enforced to avoid depleting archive resources.
To address this, several curated datasets are regularly extracted from the
archive and made available to researchers in ways suitable for mass analysis.

\subsubsection{The Software Heritage \emph{Graph} Dataset}\label{swh:graphdataset}

Consider the data model discussed in Section~\ref{swh:sec:datamodel}. The
entire archive graph is exported periodically as the \emph{Software Heritage
  Graph Dataset}~\cite{swh-msr2019-dataset}. Note the word ``graph'' in there,
which characterises this particular dataset and denotes that \emph{only} the
graph is included in the dataset, up to the content of its leave nodes,
excluded (for size reasons). This dataset is suitable for analysing source code
\emph{metadata}, including commit information, file names, software provenance,
code reuse, etc.; but not for textual analyses of archived source code, as that
is stored in graph leaves (see the blob dataset below for how to analyse actual
code).

The data model of the graph dataset is a relational representation of the
archive Merkle DAG, with one ``table'' for each type of node: blobs,
directories, commits, releases, and snapshots. Each table entry is associated
with several attributes, such as multiple checksums for blobs, file names and
attributes for directories, commit messages and timestamps for commits, etc.
The full schema is documented at
\url{https://docs.softwareheritage.org/devel/swh-dataset/graph/schema.html}.

In practical terms, the dataset is distributed as a set of Apache ORC files for
each table, suitable for loading into scale-out columnar-oriented data
processing frameworks such as Spark and Hadoop. The ORC files can be downloaded
from the public Amazon S3 bucket \url{s3://softwareheritage/graph/}. At the
time of writing the most recent dataset export has timestamp 2022-12-07 so, for
example, the first ORC files of the commit table are:

\begin{lstlisting}[language=bash,basicstyle=\ttfamily\footnotesize,frame=single]
$ aws s3 ls --no-sign-request s3://softwareheritage/graph/2022-12-07/orc/revision/
2022-12-13 17:41:44 3099338621 revision-[..]-f9492019c788.orc
2022-12-13 17:32:42 4714929458 revision-[..]-42da526d2964.orc
2022-12-13 17:57:00 3095895911 revision-[..]-9c46b558269d.orc
[..]
\end{lstlisting}

The current version of the dataset contains metadata for 13 billion source code
files, 10 billion directories, 2.7 billion commits, 35 million releases and 200 million VCS
snapshots, coming from 189 M software origins. The total size of the dataset is
11\,TiB, which makes it unpractical for use on personal machines, as opposed to
research clusters. For that reason hosted versions of the dataset are also
available on Amazon Athena and Azure Databricks. The former can be queried
using the Presto distributed SQL engine without having to download the dataset
locally. For example, the following query will return the most common first
word stems used in commit messages across more than 2.7 billion commits in just a few
seconds:

\begin{lstlisting}[language=sql,basicstyle=\ttfamily\small,frame=single,caption={Simple SQL query to get the 4 topmost commit verb stems},label={swh:lst:topcommitstems}]
SELECT count(*) as c,word FROM (
  SELECT word_stem(lower(split_part(trim(from_utf8(message)), ' ', 1))) as word
  from revision WHERE length(message) < 1000000)
WHERE word != ''
GROUP BY word ORDER BY c DESC LIMIT 4
\end{lstlisting}

For the curious reader the (unsurprising) results of the query look like
this:
\begin{center}
  \begin{tabular}{l@{~~}@{~~}l}
  \toprule
    \textbf{Count} & \textbf{Word} \\
    \midrule
    \num{294 369 196} & updat \\
    \num{178 738 450} & merg \\
    \num{152 441 261} & add \\
    \num{113 924 516} & fix \\
    \bottomrule
  \end{tabular}
\end{center}
More complex queries and examples can be found in previous
work~\cite{swh-msr2019-dataset}. For more details about using the graph dataset
we refer the reader to its technical documentation at
\url{https://docs.softwareheritage.org/devel/swh-dataset/graph/}.

In addition to the research highlights presented later in this chapter, the
Software Heritage graph dataset has been used as subject of study for the 2020
edition of the MSR (Mining Software Repositories) mining challenge, where
students and young researchers in software repository mining have used it to
solve the most interesting mining problems they could think of. To facilitate
their task ``teaser'' datasets ---data samples with exactly the same shape of the
full dataset, but much smaller--- have also been produced and can
be used by researchers to understand how the dataset works before attacking its
full scale. For example, the \texttt{popular-3k-python} teaser contains a
subset of \num{2,197} popular repositories tagged as implemented in Python and
being popular according to various metrics (e.g., GitHub stars, PyPI download
statistics, etc.). The \texttt{gitlab-all} teaser  corresponds
to all public repositories on \url{gitlab.com} (as of December 2020), an often
neglected ecosystem of Git repositories, which is interest to study to avoid
(or compare against) GitHub-specific biases.

\subsubsection{Accessing Source Code Files}

All source code files archived by Software Heritage are spread across multiple
copies and also mirrored to the public Amazon S3 bucket
\url{s3://softwareheritage/content/}. From there, individual files can be
retrieved, possibly massively and in parallel, based on their SHA1 checksums.
Starting from SWHIDs one can obtain SHA1 checksums using the \texttt{content}
table of the graph dataset and then access the associated content as follows:

\begin{lstlisting}[language=bash,basicstyle=\ttfamily\footnotesize,frame=single]
$ aws s3 cp s3://softwareheritage/content/\
    8624bcdae55baeef00cd11d5dfcfa60f68710a02 .
download: s3://softwareheritage/content/8624b[..] to ./8624b[..]

$ zcat 8624bcdae55baeef00cd11d5dfcfa60f68710a02 | sha1sum 
8624bcdae55baeef00cd11d5dfcfa60f68710a02  -

$ zcat 8624bcdae55baeef00cd11d5dfcfa60f68710a02 | head
                    GNU GENERAL PUBLIC LICENSE
                       Version 3, 29 June 2007
[..]
\end{lstlisting}

Note that individual files are gzip-compressed to further reduce storage size.

The general empirical analysis workflow involves three simple steps: identify the source code files of
interest using the metadata available in the graph dataset, obtain their
checksum identifiers, and then retrieve them in batch and in parallel from
public cloud providers. This process scales well up to many million files to be
analysed. For even larger-scale experiments, \eg analysing \emph{all} source
code files archived at Software Heritage, research institutions may consider
setting up a local mirror of the archive.\footnote{See
  \url{https://www.softwareheritage.org/mirrors/} for details, including
  storage requirements. At the time of writing a full mirror of the archive
  requires about 1\,PiB of raw storage.}

\subsubsection{License Dataset}

In addition to datasets that correspond to the actual content of the archive,
i.e., source code artifacts as encountered among public code, it is also
possible to curate \emph{derived} datasets extracted from Software Heritage for
the specific use cases or fields of endeavours.

As of today one notable example of such a derived dataset is the \emph{license
  blob dataset}, available at
\url{https://annex.softwareheritage.org/public/dataset/license-blobs/} and
described in~\cite{msr-2022-foss-licenses}. It consists of the largest known
dataset of the complete texts of free/open source software (FOSS) license
variants. To assemble it the authors collected from the Software Heritage
archive all versions of files whose names are commonly used to convey licensing
terms to software users and developers, e.g., \texttt{COPYRIGHT},
\texttt{LICENSE}, etc.~(the exact pattern is documented as part of the dataset
replication package).

The dataset consists of 6.5 million unique license files that can be used to
conduct empirical studies on open source licensing, training of automated
license classifiers, natural language processing (NLP) analyses of legal texts,
as well as historical and phylogenetic studies on FOSS licensing. Additional
metadata about shipped license files are also provided, making the dataset
ready to use in various empirical software engineering contexts. Metadata
include: file length measures, detected MIME type, detected
SPDX~\cite{stewart2010spdxspec} license (using
ScanCode~\cite{scancode-toolkit}, a state-of-the-art tool for license
detection), example origin (e.g., GitHub repository), oldest public commit in
which the license appeared. The dataset is released as open data as an archive
file containing all deduplicated license files, plus several portable CSV files
for metadata, referencing files via cryptographic checksums.

\section{Research Highlights}
\label{SWH:sec:research}

The datasets discussed in the previous section have been used to tackle
research problems in empirical software engineering and neighbouring
fields. In this section we provide brief highlights on the most interesting of
them.

\subsection{Enabling Artifact Access and (Large-Scale) Analysis}

Applied research in various fields has been conducted to ease access to such a
huge amount of data as the Software Heritage archive for empirical researchers.
This kind of research is not, strictly speaking, research \emph{enabled by} the
availability of the archive to solve software engineering problems, but rather
research \emph{motivated by} the practical need of empowering fellow scholars
to do so empirically.

As a first example \emph{SwhFS (the ``Software Heritage File
  System'')}~\cite{swh-fuse-icse2021} is a virtual filesystem developed using
the Linux FUSE (Filesystem in User SpacE) framework that can ``mount'', in the
UNIX tradition, selected parts of the archive as if they were available locally
as part of your filesystem. For example, starting from a known SWHID, one can
for instance:

\begin{lstlisting}[language=bash,basicstyle=\ttfamily\small,frame=single]
$ mkdir swhfs
$ swh fs mount swhfs/  # mount the archive
$ cd swhfs/

$ cat archive/swh:1:cnt:c839dea9e8e6f0528b468214348fee8669b305b2
#include <stdio.h>

int main(void) {
    printf("Hello, World!\n");
}

$ cd archive/swh:1:dir:1fee702c7e6d14395bbf\
5ac3598e73bcbf97b030
$ ls | wc -l
127
$ grep -i antenna THE_LUNAR_LANDING.s | cut -f 5
# IS THE LR ANTENNA IN POSITION 1 YET
# BRANCH IF ANTENNA ALREADY IN POSITION 1
\end{lstlisting}

In the second example we are grepping through the code of Apollo 11 guidance
computer code, searching for reference to antennas.

SwhFS allows to bridge the gap between classic UNIX-like mining tools, which are
often relied upon in the  fields of empirical software engineering and software repository mining, as well as by the Software Heritage
APIs. However, it is not suitable for very large scale mining, due to the fact
that seemingly local archive access pass through the public Internet (with
caching, but still not suitable for large experiments).

\emph{swh-graph}~\cite{saner-2020-swh-graph} is a way to enable such
large-scale experiments. The main idea behind its approach is to adapt and
apply \emph{graph compression} techniques, commonly used for graphs such as the
Web or social network, to the Merkle DAG graph that underpins the Software
Heritage archive. The main research question addressed by swh-graph is:
\begin{quotation}
  Is it possible to efficiently perform software development history analyses
  at ultra-large scale, on a single, relatively cheap machine?
\end{quotation}
The answer is affirmative. As of today the entire structure of the Software
Heritage graph ($\approx$25\,billion nodes + 350\,billion edges) can be loaded in memory on
a single machine equipped with $\approx\,$200\,GiB of RAM (roughly: 100\,GiB
for the direct graph + 100\,GiB for its transposed version, which is useful in
many research use cases such as source code provenance analysis). While
significant and not suitable for personal machines, such requirements are
perfectly fine for server-grade hardware on the market, with an investment of a
few thousand US dollars in RAM. Once loaded the entire graph can be visited in
full in just a few hours and a single path visit from end-to-end can be
performed in tens of nanoseconds per edge, close to the cost of a single memory
access per edge.

In practical terms, this allows to answer queries such as ``where does this
file/directory/commit come from'' or ``list the entire content of this
repositories'' in fractions of seconds (depending just on the size of the
answer, in most cases) fully in memory, without having to rely on a DBMS or
even just disk accesses. The price to pay for this is that: (1) the compressed
graph representation loaded in memory is derived from the main archive and not
incremental (it should periodically be recreated) and (2) only the graph
structure and selected metadata fit in RAM, others reside on disk (although
using compressed representations as well~\cite{pietri2021phd}) and need to be
memory mapped for efficient access to frequently accessed information.

Finally, the archive also provides interesting use cases for database research.
Recently, Wellenzohn \etal~\cite{vldb-2022-rscas-swh} has used it to develop a
novel type of \emph{content-and-structure (CAS) index}, capable of indexing
over time the evolution of properties associated to specific graph nodes, e.g.,
a file content residing at a given place in a repository changing over time
together with its metadata (last modified timestamp, author, etc.). While these
indexes existed before, their deployment and efficient
pre-population were still unexplored at this scale.

\subsection{Software Provenance and Evolution}

The peculiar structure---a fully deduplicated Merkle DAG---and
comprehensiveness of the Software Heritage archive provides a powerful
observation point and tool on the evolution and provenance of public source
code artifacts. In particular it is possible, on the one hand, to navigate the
Merkle DAG \emph{backwards}, starting from any artifact of interest (source code
file, directory, commit, etc.), to obtain the full list of all places (e.g.,
different repositories) where it has ever been distributed from. This area is
referred to as \emph{software provenance} and, in its simplest form, deals with
determining the \emph{original} (i.e., earliest) distribution place of a given
artifact. More generally, being able to identify \emph{all} places that have
ever distributed it provides a way to measure software impact, track out of
date copies or clones, and more.

Rousseau \etal~\cite{swh-provenance-emse} used the Software Heritage archive in a
study that made two relevant contributions in this
area. First, exploiting the fact that commits are deduplicated and
timestamped, they verified that \emph{the growth of public code} as a whole, at
least as it is observable from the lenses of Software Heritage \emph{is
  exponential}: the amount of original commits (i.e., commits never observed
before throughout the archive, no matter the origin repository) in public
source code doubles every $\approx$\,30 months and has been doing so for the
past 20 years. If, on the other hand, we look at original source code blobs
(i.e., files whose content has never been observed before throughout the
archive, up to that point in time), the overall trends remains the same, only
the speed changes: the amount of original public source code blobs doubles
every $\approx$\,22 months. These are remarkable findings for software
evolution, which had never been verified before at this macro-level.

Second, the authors showed how to model software provenance compactly, so that it can
be represented (space-)efficiently at the scale of Software Heritage, and can
be used to address software audit use cases which are commonplace in open
source compliance scenarios, merger and acquisition audits, etc.

\subsection{Software Forks}

The same characteristics that enable studying the evolution and provenance of
public code artifacts can be leveraged to study the global ecosystem of
software forks. In particular, the fact that commits are fully deduplicated
allows to detect forks---both collaborative ones, such as those created on
social coding platforms to submit pull requests, and hostile ones used to bring
the project in a different direction---even when they are not created on the
same platform. It is possible to detect the fork of a project originally
created on GitHub and living on GitLab.com, or vice-versa, based on the fact
that the respective repositories share a common commit history.

This is important as a methodological point for empirical researchers, because
by relying only on platform metadata (e.g., the fact that a repository
\emph{has been created} by clicking on a ``fork'' button on the GitHub user
interface) researchers risk overlooking other relevant forks.  In previous
work Zacchiroli~\cite{msr-2022-foss-licenses} provided a classification of the
type of forks based on whether they are explicitly tracked as being forks of one
another on a coding platform (Type 1 forks), they share at least one commit
(Type 2), or they share a common root directory at some point in their
histories (Type 3). He empirically verified that between $3.8\%$ and $16\%$
forks could be overlooked by considering only type 1 forks, possibly inducing a
significant threat to validity for empirical analyses of forks that strive to
be comprehensive.

Along the same lines, Bhattacharjee \etal~\cite{bhattacharjee2020crossplatformforks}
(participants in the MSR 2020 mining challenge) focus their analyses on
``cross-platform'' forks between GitHub and GitLab.com, identifying several
cases in which interesting development activity can be found on GitLab even for
projects initially mirrored from GitHub.

\subsection{Diversity, Equity, and Inclusion}

Diversity, equity, and inclusion studies (DE\&I) are hot research topics in the
area of human aspects of software engineering. Free/open source software
artifacts, as archived by Software Heritage, provides a wealth of data for
analysing evolutionary DE\&I trends, in particular in the very long term and at
the largest scale attempted thus far.

A recent study by Zacchiroli~\cite{ieee-sw-gender-swh} has used Software
Heritage to explore the trend of \emph{gender diversity} over a time period of
50 years. He conducted a longitudinal study of the population of
contributors to publicly available software source code, analysing 1.6 billion
commits corresponding to the development history of 120 million projects,
contributed by 33 million distinct authors over a period of 50 years. At this
scale authors cannot be interviewed to ask their gender, nor cross-checking
with large-enough complementary dataset was possible. Instead, automated
detection based on census data from around the world and the gender-guesser
tool (benchmarked for accuracy, and popular in the field) was used. Results
show that, while the amount of commits by female authors remains very low
overall (male authors have contributed more than 92\% of public code commits
over the 50 years leading to 2019), there is evidence of a stable long-term
increase in their proportion over all contributions (with the ratio of commits
by female authors growing steadily over 15 years, reaching in 2019 for the
first time 10\% of all contributions to public code).

Follow up studies have added the spatial dimension investigating the
\emph{geographic gap} in addition to the gender one. Rossi \etal~\cite{msr-2022-foss-geography}
have developed techniques to detect the
geographic origin of authors of Software Heritage commit, using as signals the
timezone offset and the author names (compared against census date from around
the world). Results over 50 years of development history show evidence of the
early dominance of North America in open source software, later joined by
Europe. After that period, the geographic diversity in public code has been
constantly increasing, with more and more contributions coming from Central and
South Asia (comprising India), Russia, Africa, Central and South America.

Finally, Rossi \etal~\cite{icse-seis-2022-gender} put together the temporal
and spatial dimension using the Software Heritage archive to investigate whether the
ratio of women participation over time shows notable differences around the
world, at the granularity of 20 macro regions. The main result is that the
increased trend of women participation is indeed a world-wide phenomenon, with
the exception of specific regions of Asia where the increase is either slowed
or completely flat. An incidental finding is also worth noting: the positive
trend of increased women participation observed up to 2019 has been reversed by
the COVID-19 pandemic, with the \emph{ratio} of both contributions by and
active female authors decreasing sharply starting at about that time.

These studies show how social aspects of software engineering can benefit from
large-scale empirical studies and how they can be enabled by comprehensive,
public archives of public code artifacts.

\section{Building the Software Pillar of Open Science}
\label{SWH:sec:openscience}

\begin{quote}
  Software plays a key role in scientific research, and it can be a tool, a result, and a research object. [...] France will support the development and preservation of source code – inseparable from the support of humanity’s technical and scientific knowledge – and it will, from this position, continue its support for the Software Heritage universal archive. So as to create an ecosystem that connects code, data and publications, the collaboration between the national open archive HAL, the national research data platform Recherche Data Gouv, the scientific publishing sector and Software Heritage will be strengthened.\\
  \mbox{}\hfill \emph{Second french national plan for open science, July 2021}~\cite{swh-PNSO2}
\end{quote}

Software is \emph{an essential research output}, and its source code implements
and describes data generation and collection, data visualisation, data analysis,
data transformation, and data processing with a level of precision that is not
met by scholarly articles alone. Publicly accessible software source code allows
a better understanding of the process that leads to research results, and open
source software allows researchers to build upon the results obtained by others,
provided proper mechanisms are put in place to make sure that software source
code is preserved and that it is referenced in a persistent way.

There is a growing general awareness of its importance for supporting the
research process~\cite{Borgman2012, Stodden-reprod-2012, Hinsen2013}.  Many
research communities focus on the issue of \emph{scientific reproducibility}
and strongly encourage making the source code of the artefact available by
archiving it in publicly accessible long-term archives; some have even put in
place mechanisms to assess research software, like the \emph{Artefact Evaluation} 
process introduced in the ESEC-FSE 2011 conference
and now widely adopted by many computer science
conferences~\cite{Dagstuhl-Artefacts-2016},
and the ACM \emph{Artifact Review and Badging} program.~\footnote{\url{https://www.acm.org/publications/policies/artifact-review-badging}}
 Other raise the complementary
issues of making it easier to discover existing research software, and giving
academic credit to its authors~\cite{SoftwareCitationPrinciples-2016,
  HowisonBullard2016,Lamprecht2019}.

These important issues are similar in spirit to those that led to the now
popular FAIR data movement~\cite{FAIRdefinition2016}, and as a first step it is
important to clearly identify the different concerns that come into play when
addressing software, and in particular its source code, as a research output.
They can be classified as follows:
\begin{description}
\item[\itshape Archival:] software artifacts must be properly
  \textbf{archived}, to ensure we can \emph{retrieve} them at a later time;
\item[\itshape Reference:] software artifacts must be properly
  \textbf{referenced} to ensure we can \emph{identify} the exact code, among
  many potentially archived copies, used for reproducing a specific experiment;
\item[\itshape Description:] software artifacts must be equipped with proper
  \textbf{metadata} to make it easy to \emph{find} them in a catalog or through
  a search engine;
\item[\itshape Citation:] research software must be properly \textbf{cited} in
  research articles in order to give \emph{credit} to the people that
  contributed to it.
\end{description}

These are not only different
concerns, but also \emph{separate} ones. Establishing proper \emph{credit} for
contributors via \emph{citations} or providing proper metadata to
\emph{describe} the artifacts requires a \emph{curation} process~\cite{swmath,
  ASCL, swh-hal-02475835} and is way more complex than simply providing
stable, intrinsic identifiers to \emph{reference} a precise version of a
software source code for reproducibility purposes~\cite{HowisonBullard2016,
  2020GtCitation, cise-2020-doi}.  Also, as remarked in~\cite{Hinsen2013,
  2020GtCitation}, research software is often a thin layer on top of a large
number of software dependencies that are developed and maintained outside of
academia, so the usual approach based on institutional archives is not
sufficient to cover all the software that is relevant for reproducibility of
research.

In this section, we focus on the first two concerns, \emph{archival} and
\emph{reference}, that can be addressed fully by leveraging the Software
Heritage archive, but we also describe how Software Heritage contributes
through its ecosystem to the two other concerns.

\subsection{Software in the Scholarly Ecosystem}

Presenting results in journal or conference articles has always been part of the
research activity. The growing trend, however, is to include software to support
or demonstrate such results. This activity can be a significant part of academic
work and must be properly taken into account when researchers are
evaluated~\cite{2020GtCitation,SoftwareCitationPrinciples-2016}.

\begin{figure*}[thb]
  \includegraphics[width=\textwidth]{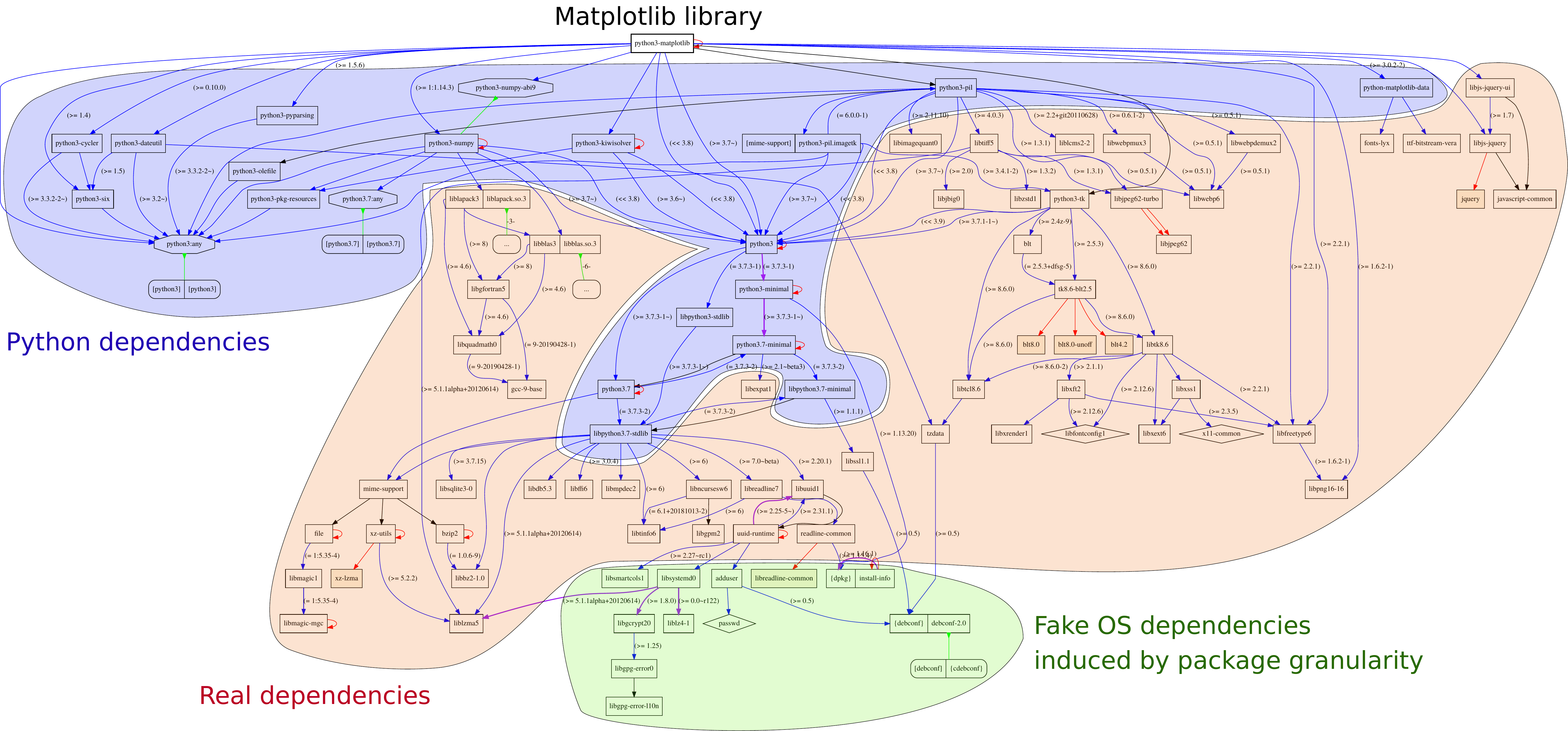}
  \caption{Direct and indirect
    dependencies for a specific python package (matplotlib). In blue the
    Python dependencies, in red the ``true'' system dependencies
    incurred by python (e.g., the \texttt{libc} or
    \texttt{libjpeg62}), in green some dependencies triggered by the
    package management system but which are very likely not used by python
    (e.g., \texttt{adduser} or \texttt{dpkg}).}
    \label{swh:fig:matplotlib}
\end{figure*}

Software source code developed by researchers is only \emph{a thin layer} on top
of the complex web of software components, most of them developed outside of
academia, that are necessary to produce scientific results: as an example,
Figure ~\ref{swh:fig:matplotlib} shows the many components that are needed by
the popular \verb|matplotlib| library~\cite{Hunter:2007}.

As a consequence, scholarly infrastructures that support software source code
written in academia must go the extra mile to ensure they adopt standards and
provide mechanisms that are compatible with the ones used by tens of millions of
non-academic software developers worldwide. They also need to ensure that the
large amount of software components that are developed outside academia, but are
relevant for research activities, are properly taken into account.

Over the recent years, there have been a number of initiatives to add support
for software artifacts in the scholarly world, that fall short of satisfying
these requirements. They can be roughly classified in two categories:

\begin{description}
  \item[{\bf overlays on public forges}] provide links from articles to the source code repository of the associated software artifact as found on a public code hosting platform (forge); typical examples are websites like \url{https://paperswithcode.com/}, \url{http://www.replicabilitystamp.org/} and the \emph{Code and data} links recently introduced in \verb|ArXiv.org|.
  
  \item[{\bf deposits in academic repositories}] take snapshots of a given state of the source code, usually in the form of a \verb|.zip| or \verb|.tar| file, and store it in the repository exactly like an article or a dataset, with an associated publisher identifier; typical examples in Computer Science is the ACM Digital Library, but there are a number of general academic repositories where software artefacts have been deposited, like FigShare and Zenodo.
\end{description}

The approaches in the first category rely on code hosting platforms that do not guarantee \emph{persistence} of the software artifact: the author of a project may alter, rename, or remove it, and we have seen that code hosting platforms can be discontinued, or decide to remove large amount of projects.\footnote{Google Code and Gitorious.org were shut down in 2015, Bitbucket removed support for the Mercurial VCS in 2020, and in 2022 Gitlab.com considered removing all projects inactive for more than a year.}

The approaches in the second category do take into account persistence, as they archive software snapshots, but they loose the \emph{version control history} and do not provide the \emph{granularity} needed to reference the internal components of a software artifact (directories, files, snippets).

And none of the initiatives in these categories provides a means to properly archive and reference the numerous external dependencies of software artefacts.

This is where Software Heritage comes into play for Open Science, by providing an \emph{archive designed for software} that provides persistence, preserves the version control history, supports granularity in the identification of software artefacts and their components, and harvests all publicly available source code.

The differences described above are summarised in the following table, where we only consider infrastructures in the second category described above, as they are the only one assuming the mission to archive their contents. We also take into account additional features found in academic repositories, like the possibility of depositing content with an \emph{embargo} period, which is not possible on Software Heritage, and the existence of a curation process to obtain qualified metadata, which is currently out of scope of Software Heritage.

\begin{table}[h]
\centering
\begin{center}
\caption{Comparison of infrastructures for archiving research software. The various granularities of identifiers are abbreviated with the same convention used in SWHIDs (\emph{snp} for snapshot, etc.), plus the abbreviation \emph{frg} that stands for the ability to identify a \emph{code fragment}.}
\begin{tabular}{c|c|c|c|c|c}
\toprule
\diagbox{Criteria}{Infrastructure}
                                 & Software Heritage          & ACM DL      & HAL         & Figshare    & Zenodo    \\
\midrule
\multirow{3}{*}{identifier}      &               &            & extrinsic   &    &             \\
                                 & intrinsic     & extrinsic  & \emph{+ intrinsic} & extrinsic & extrinsic  \\  
                                 &               &            & \emph{(via SWH)} &             &            \\  
&&&&&\\
\multirow{2}{*}{granularity}     & snp, rel, rev     &            &             &             &            \\
                                 & dir, cnt, frg     & dir        & dir         & dir         & rel, dir   \\
&&&&&\\
\multirow{3}{*}{archival}        & harvest       &            &             &             &            \\
                                 & deposit       & deposit    & deposit     & deposit     & deposit    \\
                                 & save code now &            &             &             &            \\
&&&&&\\
history                          & full VCS      & no         & no          & no          & releases   \\
&&&&&\\
browse code                      & yes           & no          & no          & no          & no        \\
&&&&&\\
scope                            & universal     & discipline  & academic    & academic    & academic  \\
&&&&&\\
embargo                          & no            & no          & yes         & yes         & yes       \\
&&&&&\\
curation                         & no            & yes         & yes         & no          & no        \\
&&&&&\\
\multirow{5}{*}{integration}     & BitBucket,    &             &             &             &           \\
                                 & SourceForge,  &             &             &             &           \\
                                 & GitHub,       &             & SWH         &             & GitHub    \\
                                 & Gitea, Gitlab, &             &             &             &           \\
                                 & HAL, etc.     &             &             &             &           \\
\bottomrule
\end{tabular}
\end{center}
\end{table}

\subsection{Extending the Scholarly Ecosystem Architecture to Software}

In the framework of the European Open Science Cloud initiative (EOSC), a
working group has been tasked in 2019 to bring together representatives from a
broad spectrum of scholarly infrastructures to study these issues and propose
concrete ways to address theme. The result, known as the EOSC Scholarly Infrastructures for Research Software (SIRS)
report~\cite{swh-SIRSReport2020} was published in 2020 and provides a detailed
analysis of the existing infrastructures, their relationships, and the
workflows that are needed to properly support software as a research result on
par with publications and data.

\begin{figure*}[thb]
  \begin{center}
    \includegraphics[width=.8\textwidth]{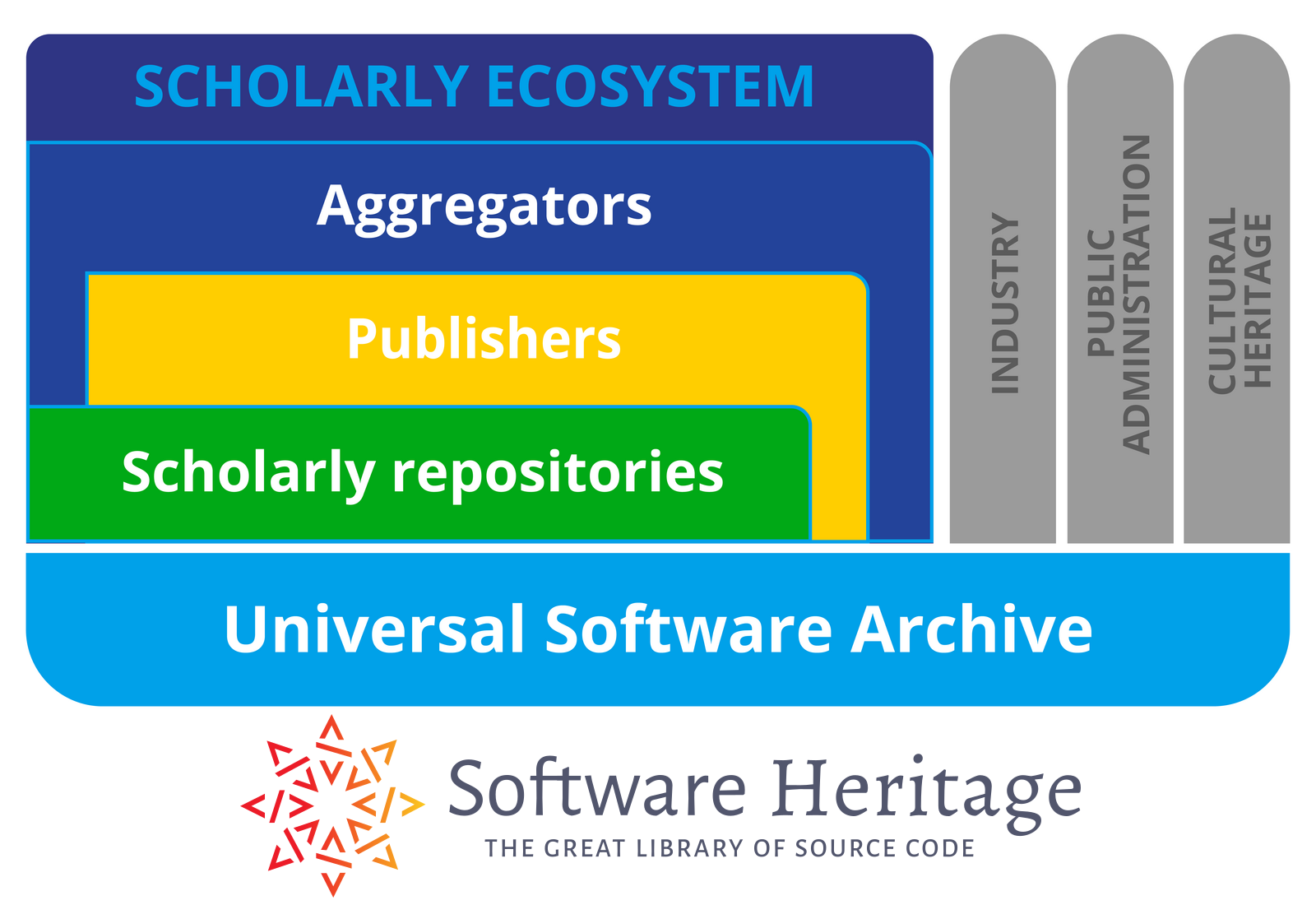}
  \end{center}
    \caption{Overview of the high level architecture of scholarly infrastructures for research software, as described in the EOSC SIRS report}
    \label{swh:fig:sirsarchitecture}
\end{figure*}

Figure~\ref{swh:fig:sirsarchitecture} presents the main categories of identified actors:
\begin{description}
\item[\itshape Scholarly repositories:] services that have as one of their
  primary goals the long-term preservation of the digital content that they
  collect.
\item[\itshape Academic publishers:] organisations that prepare submitted
  research texts, possibly with associated source code and data, to produce a
  publication and manage the dissemination, promotion, and archival
  process. Software and data can be part of the main publication, or assets
  given as supplementary materials depending on the policy of the journal.
\item[\itshape Aggregators:] services that collect information about digital
  content from a variety of sources with the primary goal of increasing its
  discoverability, and possibly adding value to this information via processes
  like curation, abstraction, classification, and linking.
\end{description}

These actors have a long history of collaboration around research articles, with
well defined workflows and collaborations. The novelty here is the fact that to
handle research software, it is no longer possible to work in isolation inside
the academic world, for the reasons explained previously: one needs a means to
share information and work with other ecosystems where software is present, like
industry and public administration.

One key finding of the EOSC SIRS Report is that Software Heritage provides the
shared basic architectural layer that allows to interconnect all these
ecosystems, because of its unified approach to archiving and referencing all
software artefacts, independently of the tools or platforms used to develop
or distribute the software involved.

\subsection{Growing Technical and Policy Support}

In order to take advantage of the services provide by Software Heritage in this
setting, a broad spectrum of actions have been started, and are ongoing.
We briefly survey here the ones that are most relevant at the time of writing.

At the national level, France has developed a multi-annual plan on Open Science
that includes research software~\cite{swh-PNSO,swh-PNSO2}, and consistently implemented
this plan through a series of steps that range from technical development to
policy measures.

\begin{figure*}[thb]
  \begin{center}
    \includegraphics[width=.8\textwidth]{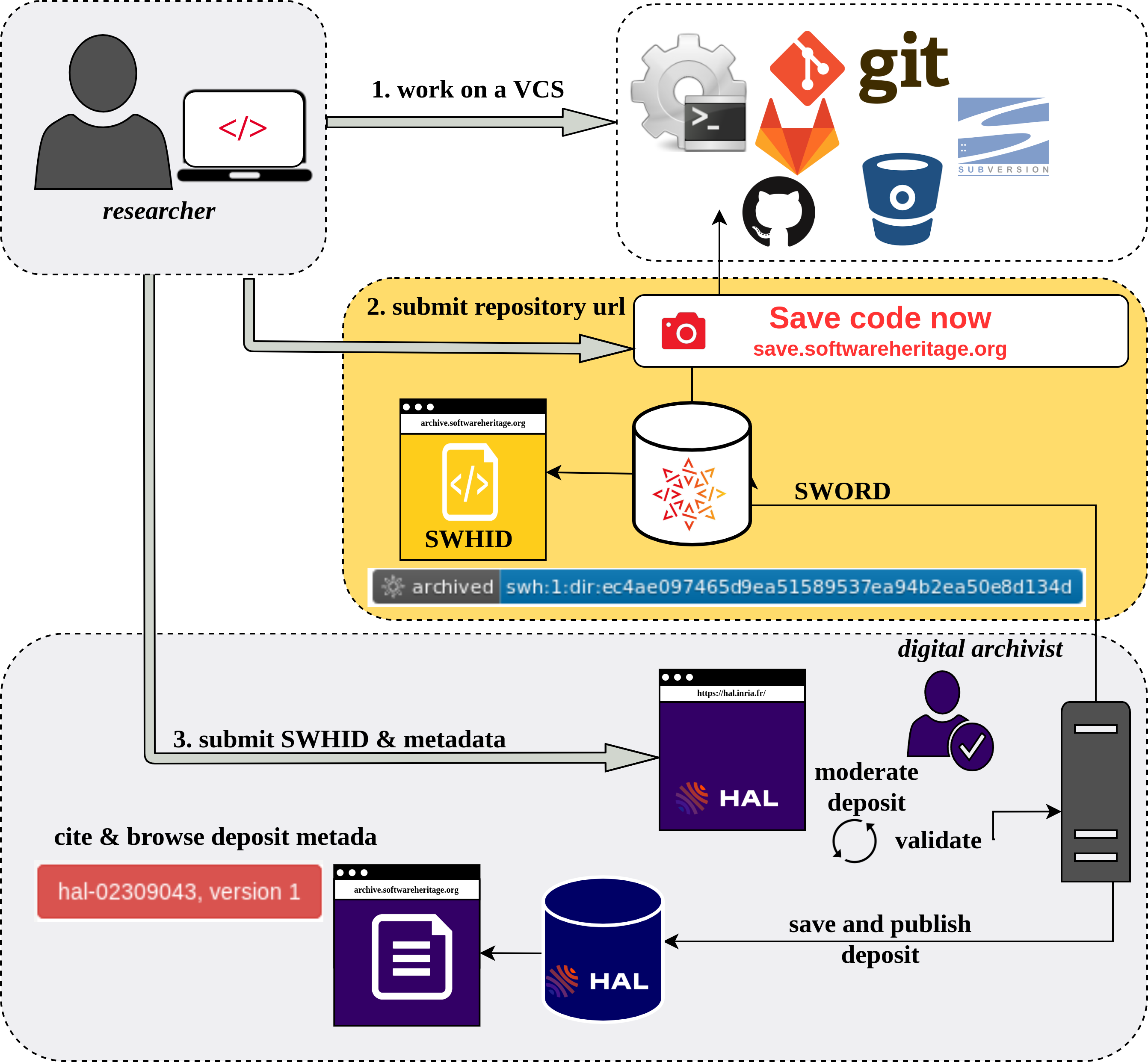}
  \end{center}
  \caption{Overview of the interplay between HAL and Software Heritage for research software}
    \label{swh:fig:halswh}
\end{figure*}

On the technical side, the French national open access repository HAL~\cite{swh-hal-02475835}
(analogous to the popular arXiv service\footnote{\url{https://arxiv.org}}) has been integrated with the Software Heritage
archive. The integration allows researchers to have their software projects
archived and referenced in Software Heritage, while curated rich metadata and
citation information is made available on HAL~\cite{swh-hal-02475835}, with a streamlined process
depicted in Figure~\ref{swh:fig:halswh}.

On the policy side, the second french national plan for open
science~\cite{swh-PNSO2}, published in July 2021, prescribes the use of
Software Heritage and HAL for all the research software produced in France, and
Software Heritage is now listed in the official national roadmap of research
infrastructures published in February 2022~\cite{swh-2022-French-roadmap}.

This approach is now being pushed forward at the European level, through funding
for consortia that will build the needed connectors between Software Heritage
and several infrastructures and technologies used in academia, using the French
experience as a reference.  Most notably, the FAIRCORE4EOSC~\cite{swh-fc4e}
European project include plans to build connectors with scholarly repository
systems like Dataverse~\cite{swh-dataverse} and InvenioRDM~\cite{swh-inveniordm}
(the white-label variant of Zenodo), publishers like
Dagstuhl~\cite{swh-dagstuhl} and Episcience~\cite{swh-episciences}, and
aggregators like swMath~\cite{swh-swmath} and OpenAire~\cite{web:openaire}.

\subsection{Supporting Researchers}

The growing awareness about the importance of software as a research output will
inevitably bring new recommendations for research activity, that will eventually
become obligations for researchers, as we have seen with publications and data.

Through the collaboration with academic infrastructures, Software Heritage is
striving to develop mechanisms that minimise the extra burden for researchers,
and we mention here a few examples.

A newly released extension, codename \verb|updateswh|, for the popular web
browsers Firefox and Google Chrome allows to trigger archival in just one click
for any public repository hosted on BitBucket, GitLab (.com, and any instance),
GitHub and any instance of Gitea. It also allows to access in one click the
archived version of the repository and obtain the associated SWHID identifier.

Integration with web hooks is available for a variety of code hosting platforms,
including BitBucket, GitHub, GitLab.com and Source forge, as well as for
instances of GitLab and Gitea, which enable owners of projects hosted on those
platforms to trigger archival automatically on any new release, reducing the
burden on researchers even more.

Software Heritage will try to detect and parse intrinsic metadata present in
software projects independently of the format chosen, but we see the value of
standardising on a common format. This is why, with all academic platforms we
are working with, we are advocating the use of \verb|codemeta.json|, a machine
readable file based on the CodeMeta extension of \verb|schema.org|, to retrieve
automatically metadata associated to software artifact, in order to avoid the
need for researchers to fill forms when declaring software artifacts in academic
catalogs, following the schema put in place with the HAL national open access
portal.

Finally, we have released the \verb|biblatex-software| bibliographic style
extension to make it easy to cite software artefacts in publications written
using the popular \LaTeX\ framework.

\section{Conclusions and Perspectives}
\label{SWH:sec:closing}

In conclusion, the Software Heritage ecosystem is a useful resource for both
software engineering studies and for Open Science. As an infrastructure for
research on software engineering, the archive provides numerous benefits. The
SWHID intrinsic identifiers make it easier for researchers to identify and track
software artifacts across different repositories and systems. The uniform data
structure used by the archive abstracts away all the details of software forges
and package managers, providing a standardised representation of software code
that is easy to use and analyse. The availability of the open datasets makes it
possible to tailor experiments to one's needs and improves their
reproducibility. An obvious direction at the time of writing is to leverage
Software Heritage's extensive source code corpus for pre-training large language
models.  Future collaborations may lead to integrate functionalities like the
domain-specific language from the Boa project or the efficient data structures
of the World of Code project, enabling researchers to run more specialised
queries and achieve more detailed insights.

Regarding the Open Science aspect, Software Heritage already offers the
reference archive for all publicly available research software. The next step is
to interconnect it with a growing number of scholarly infrastructures, which
will increase reproducibility of research in all fields, and support software
citation directly from the archive, contributing to increasing visibility of
research software.

Going forward, we believe that Software Heritage will provide a unique
observatory for the whole software development ecosystem, both in academia and
outside of it. We hope that with growing adoption it will play an increasingly
valuable role in advancing the state of software engineering research and in
supporting the software pillar of open science.

\bibliographystyle{spmpsci}
\bibliography{references.bib}


\backmatter
\end{document}